\documentclass[twocolumn,showpacs,preprintnumbers,amsmath,amssymb,superscriptaddress]{revtex4}
\usepackage[utf8]{inputenc}
\usepackage{longtable}
\usepackage{color}
\usepackage{graphicx}
\usepackage{dcolumn}
\usepackage{bm}
\begin{document}
\title {Direct Detection of MeV-Scale Dark Matter Utilizing Germanium Internal Amplification for the Charge Created by the Ionization of Impurities}
\newcommand{\usd}{Department of Physics, The University of South Dakota, Vermillion, South Dakota 57069}
\newcommand{\yru}{College of Physics, Yangtz River University, Jingzhou 434023, China}
\newcommand{\dmm}{Dongming.Mei@usd.edu}
\author{D.-M. Mei} \altaffiliation[Corresponding Author: ]{\dmm}  \affiliation{   \usd    } \affiliation{   \yru   }
\author{G.-J. Wang} \affiliation{ \usd}
\author{H. Mei} \affiliation{ \usd}
\author{G. Yang} \affiliation{ \usd} 
\author{J. Liu} \affiliation{ \usd}
\author{M. Wagner} \affiliation{ \usd}
\author{R. Panth} \affiliation{ \usd}
\author{K. Kooi} \affiliation{ \usd}
\author{Y.-Y. Li} \affiliation{ \usd}
\author{W.-Z. Wei} \affiliation{ \usd}
\begin{abstract}
Light, MeV-scale dark matter (DM) is an exciting DM candidate that is undetectable by current experiments. 
A germanium (Ge) detector utilizing internal charge amplification for the charge carriers created by the ionization of impurities is a promising new technology with experimental sensitivity for detecting MeV-scale DM. 
We analyze the physics mechanisms of the signal formation, charge creation, charge internal amplification, and the projected sensitivity for directly detecting MeV-scale DM particles. We present a design for a novel Ge detector at helium temperature ($\sim$4 K) enabling ionization of impurities from DM impacts. With large localized E-fields, the ionized excitations can be accelerated to kinetic energies larger than the Ge bandgap at which point they can create additional electron-hole pairs, producing intrinsic amplification to achieve an ultra-low energy threshold of $\sim$0.1 eV for detecting low-mass DM particles in the MeV scale. Correspondingly, such a Ge detector with 1 kg-year exposure will have high sensitivity to a DM-nucleon cross section of $\sim$5$\times$10$^{-45}$ cm$^{2}$ at a DM mass of $\sim$10 MeV/c$^{2}$ and a DM-electron cross section of $\sim$5$\times$10$^{-46}$cm$^{2}$ at a DM mass of $\sim$1 MeV/c$^2$.

\end{abstract}

\maketitle
\section{Introduction}
Observations from the 1930’s~\cite{fzw} have led to the contemporary and shocking revelation
that 96\% of the matter and energy in the universe neither emits nor absorbs electromagnetic 
radiation~\cite{ghi, rjg}.  Weakly Interacting Massive Particles (WIMPs)~\cite{jlf}
 constitute a popular candidate for dark matter (DM). These particles, with mass thought to be comparable to heavy nuclei, have a feeble and extremely short range
interaction with atomic nuclei. While WIMPs appear to interact with atomic nuclei very rarely, their collisions would cause atoms to recoil at a velocity in the order of a thousand times the speed of sound in air~\cite{mwg}. 

For over three decades, many experiments have conducted searches for WIMPs using various targets~\cite{cdms, cdex, cogent, cresst, coupp, damic, dama, darkside, drift, edelweiss, kim, lux, pandax, pico, supercdms, xenon10, xenon100, xenon1t, xmass, zeplin}. 
 These experiments are all sensitive to WIMPs with masses greater than a few GeV/c$^{2}$. The best sensitivity for WIMPs masses above 10 GeV/c$^{2}$, with a minimum of 7.7$\times$10$^{-47}$ cm$^{2}$ for 35 GeV/c$^{2}$ at 90\% confidence level, is given by the most recent results from XENON1T~\cite{xenon1t}. Despite great efforts have been made, WIMPs remain undetected. More experiments will soon come online~\cite{deap, clean, lz}. The LZ experiment will push the experimental sensitivity for WIMPs with masses greater than 10 GeV/c$^{2}$  very close to the boundary where neutrino induced backgrounds begin to constrain the experimental sensitivity~\cite{lgs, agu}.

In the past decade, light DM in the MeV-scale~\cite{res, ess1, cmh, gst} has risen to become an exciting 
DM candidate, even though its low mass makes it unreachable by current experiments. The detection of MeV-scale DM requires new detectors with an extremely low-energy threshold ($<$ 10 eV) since both electronic recoils and nuclear recoils
induced by MeV-scale DM are in the range of sub-eV to 100 eV~\cite{res}. Using XENON10 data, 
the XENON Collaboration
was able to set the first experimental limit on the MeV-scale DM detection~\cite{xenon10l}. 
More recently, Kadribasic et al. have proposed a method for using solid state detectors with directional sensitivity to DM interactions to detect low-mass DM~\cite{kad}.
 CRESST has achieved a threshold of 20 eV with a small prototype sapphire detector~\cite{sap}. DAMIC has claimed a
sensitivity to ionization $<$ 12 eV with Si CCDs and considered their method to be able to reach 1.2 eV~\cite{damic1}.

DM coupling to visible matter is assumed through weak and gravitational interactions~\cite{gju,pfs}. A common search channel is the elastic scattering between 
incoming DM particles and target nuclei. 
Current direct detection experiments search for nuclear recoils with the lowest accessible nuclear recoil energy being around 1 keV~\cite{cogent, cdex, lux, supercdms}. This corresponds to DM with masses greater than 6 GeV/c$^{2}$.
For MeV-scale DM, the average nuclear recoil 
energy gained from an elastic scattering is:
\begin{equation}
\label{dm1}
E_{nr} = q^{2}/2m_{N},
\end{equation}
which is the level of~\cite{res}:
\begin{equation}
\label{dm2}
  E_{nr} \simeq 1 \  eV \times (m_{\chi}/100 \ MeV)^2 (10 \ GeV/m_{N}),
  \end{equation}
where $q\sim$$m_{\chi}v$ is the
momentum transferred, $v$$\sim$10$^{-3}c$ is the DM velocity, $c$ is the speed of light, $m_{\chi}$
is the mass of DM, and $m_{N}$ is the mass of a nucleus. As can be seen, 
this nuclear recoil energy is in the range of $\sim$ 1 eV and  well below the lowest threshold
achieved in existing direct detection experiments.

On the other hand, coupling between the incoming DM and the orbital electrons is also possible~\cite{res, cmh, gst}. In this case, the total energy available in the scattering between DM and electrons
can be larger~\cite{res}:
\begin{equation}
E_{tot} \simeq m_{\chi}v^{2}/2 \simeq 50 \ eV \times (m_{\chi}/100 \ MeV). 
\end{equation}
However, it is 
still in the level of $\sim$ 50 eV if $m_{\chi}$ is 100 MeV/c$^{2}$. 
Consequently, conventional detector technology 
does not allow for the detection of DM much below the
GeV mass scale. Direct detection of MeV-scale DM requires new detectors with
threshold as low as sub-eV to maximize the capability of searches.

A promising technology for sensitivity to MeV-scale DM is a germanium (Ge) detector which utilizes internal charge amplification for the charge carriers created by the ionization of impurities.
We describe the design of a novel Ge detector that develops ionization amplification technology for Ge in which very large localized E-fields are used to accelerate ionized excitations produced by particle interaction to kinetic energies larger than the Ge bandgap at which point they can create additional electron-hole (e-h) pairs, producing internal amplification. This amplified charge signal could then be readout with standard high impedance JFET or HEMT~\cite{mjc} based charge amplifiers. Such a system would potentially be sensitive to single ionized excitations produced by DM interactions with both nuclei and electrons. In addition, purposeful doping of the Ge could lower the ionization threshold by $\sim \times$10 ($\sim$ 0.1 eV), making the detector sensitive to 100 keV DM via electronic recoils.

\section{The formation of signal}
\subsection{DM-nucleus and DM-electron elastic scattering processes}
The energy deposition between the incoming DM and a target nucleus or a bound electron through elastic scattering 
can be calculated using the standard halo model~\cite{gju, pfs} assuming the velocity
distribution of DM is approximately Maxwellian, with $v_{rms}$ governed by the gravitational binding and having a value $\simeq$ 270 $km$s$^{-1}$. The relative motion of the solar system ($v$=220 $km$s$^{-1}$)
through the DM halo is considered in the calculation. 
The energy deposition spectrum arises due to kinematics of elastic scattering. In the center-of-momentum frame, a DM particle scatters off a nucleus or a bound electron through an angle $\theta$, uniformly distributed between 0 and 180$^{o}$ for the isotropic scattering that occurs with zero-momentum transfer. The DM’s initial energy in the laboratory frame can be expressed as: $E_{i}$ = m$_{\chi}v^2$/2. The nuclear recoil energy  can be calculated as:
\begin{equation}
\label{recoil} 
E_{r} = E_{i}\times \frac{4\mu_{\chi N}^2}{m_{\chi}m_{N}}\times\frac{(1-cos\theta)}{2},
\end{equation}
where $\mu_{\chi N}$ is the DM-nucleus reduced mass, $\mu_{\chi N}$ = $\frac{m_{\chi}m_{N}}{m_{\chi}+m_{N}}$. 

When a DM particle collides directly with a bound electron, exciting it to a higher energy level or an unbound state, the calculation of electronic recoil energy is different from that of a nuclear recoil~\cite{ess1}. Since electrons are in a bound state, the electrons may have an arbitrarily high momentum (albeit with low probability). 
During a collision between a DM particle and a bound electron, the energy transferred to the electron, $\Delta E_{e}$, can be related to the momentum lost by the DM, $\vec{q}$, via energy conservation~\cite{ess1}:
\begin{equation}
\label{electron}
\Delta E_{e} = \vec{q}\cdot\vec{v}-\frac{q^2}{2\mu_{\chi N}}.
\end{equation}
If one assumes an angle, $\alpha$, between the momentum transfer ($\vec{q}$) and the velocity ($\vec{v}$) of the DM particle, $\Delta E_{e}$ can be written as:
\begin{equation}
\label{er1}
\Delta E_{e} = q v cos(\alpha) - \frac{q^2}{2\mu_{\chi N}}.
\end{equation}
Taking into account a fact that $\Delta E_{e}$ = $\frac{q^2}{2\overline{Z_{eff}}m_{e}}$, the energy transferred to electrons equals the squared momentum lost by the DM particle divided by two times of the effective mass of that atomic system, where m$_{e}$ is the mass of electron, then eq.~\ref{er1} can be rewritten as:
\begin{equation}
\label{er2}
\Delta E_{e} = \frac{2\overline{Z_{eff}}m_{e}v^2cos^2(\alpha)}{(1+\frac{\overline{Z_{eff}}m_{e}}{\mu_{\chi N}})^2},
\end{equation}
where Z$_{eff}$ is the effective number of orbital electrons (also called effective nuclear charge) that participate in the DM-electron scattering, which would interact with the entire atomic system. The value of Z$_{eff}$ corresponding to the electron configuration, 1s$^2$2s$^2$2p$^6$3s$^2$3p$^6$3d$^{10}$4s$^2$4p$^2$,  is given by  Clementi et al.~\cite{cle1,cle2}, as shown in Table~\ref{zeff}.
\begin{table}[htb]
\caption{Effective nuclear charges for Ge electron configuration. }
\centering
\begin{tabular}{|c|c|}
\hline
  Electron Configuration & Z$_{eff}$\\
	\hline
  1s & 31.294\\ 
  2s & 23.365 \\
  2p & 28.082 \\
  3s & 17.790 \\
  3p & 17.014 \\
  4s & 8.044 \\
  3d & 16.251 \\
  4p & 6.780 \\
	\hline
\end{tabular}
\label{zeff}
\end{table}
Utilizing the electron configuration of Ge and the values of Z$_{eff}$ in Table~\ref{zeff}, the average value of $\overline{Z_{eff}}$ is determined to be 18.989 for a Ge atom. 
Since an arbitrary-size momentum transfer is possible, the largest allowed energy transfer is found to be:
\begin{equation}
\label{electron1}
\Delta E_{e} \leq \frac{1}{2}m_{\chi}v^2.
\end{equation}
The likelihood of actually obtaining a large enough $q$ to excite an electron depends on the effective atomic number, $\overline{Z_{eff}}$, and the incident angle of a DM particle. 

\begin{figure}
\includegraphics[width=0.48\textwidth]{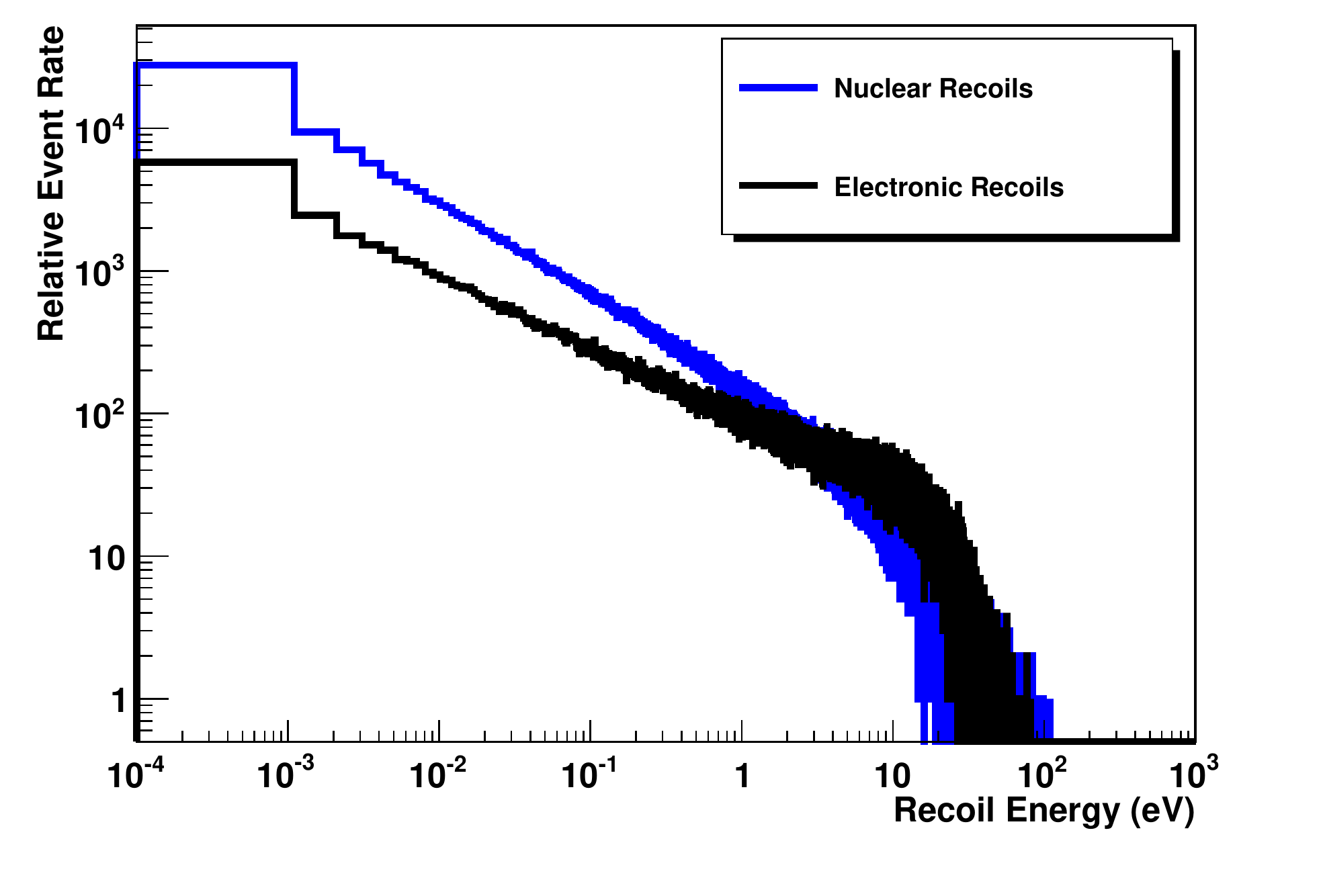}
\caption{
The relative event rate as a function of recoil energy for DM with masses 
between 0.1 MeV/c$^{2}$ to 1 GeV/c$^{2}$. 
}
\label{fig:fullSpe}
\end{figure}

With the standard halo model described above, the energy deposition is simulated as shown in
Figure~\ref{fig:fullSpe}, which shows the distributions of nuclear recoil energies induced by DM-nucleus scattering and electronic recoil energies created by DM-electron scattering with DM masses from 0.1 MeV/$c^{2}$ to 1 GeV/$c^{2}$. 
It is clear that the energy deposited by nuclear recoils is mainly in the range of sub-eV to $\sim$100 eV. 
The dissipation of such a small amount of energy in Ge is largely through the emission of phonons.
DM interacting with electrons can lead to visible signals of $\sim$ 10 eV through the following
channels: electron ionization and electronic excitation. 
Similar to DM-nucleus scattering, the energy deposited by electronic recoils is also in the range of sub-eV to $\sim$100 eV.
Such a small amount of energy is again largely dissipated through the emission of phonons. Therefore, the detection of those phonons is a major consideration for the design of the next generation
of Ge detectors. 
\subsection{The form of the detectable signature}
The bandgap energy of Ge is 0.67 eV at room temperature~\cite{ras}. It increases
slightly as temperature decreases~\cite{ypv}. The band structure of Ge is an
indirect bandgap, which means that the minimum of conduction band and the maximum of valance band 
lie at different momentum, $k$, values. When an e-h pair is created in Ge, phonons
must be involved to conserve momentum~\cite{wei}. Therefore, the average energy expended
per e-h pair in Ge at 77 K is $\sim$3 eV, much higher than the bandgap energy of 0.73 eV~\cite{wei} at the
same temperature.

Since the energy dissipation, induced by DM interactions with a nucleus or an electron, is mainly released
through the emission of phonons, the energy of phonons (E$_{phonon}$), can be estimated through E$_{phonon}=hv_{s}/a$, where $h$ is the Planck constant, $v_{s}$ is the speed of sound in Ge, and $a$ (0.565 nm) is the lattice constant of Ge. 
The value of the speed of sound in Ge depends on the polarization of phonons and the orientation of Ge crystal. For a [100] Ge crystal, $v_{s}$ = 5.4$\times$10$^{3}$m/s~\cite{gvs} for longitudinal acoustic (LA) phonons and $v_{s}$ = 3.58$\times$10$^{3}$m/s for transverse acoustic (TA) phonons~\cite{mem}. Therefore, E$_{phonon}$ can be 0.037 eV for LA phonons and 0.026 eV for TA phonons, respectively. These values agree with the early measurements made by Brockhouse and Iyengar~\cite{bro} and Others~\cite{oth}. There are also measured phonons in longitudinal optical (LO) and transverse optical (TO) bench with energies up to 0.063 eV~\cite{bro}. 
The energies of LA, TA, LO, and TO phonons are much less than 3 eV required to generate an e-h 
pair at 77 K.
Those phonons are not capable of generating e-h pairs through 
excitation of Ge atoms. 
Indirect detection of those phonons has been demonstrated
by CDMS~\cite{cdms}, EDELWEISS~\cite{edelweiss} and SuperCDMS~\cite{supercdms} with threshold energy as low as $\sim$50 eV. 

To access the large portion of the recoil energy spectra shown in Figure~\ref{fig:fullSpe}, a threshold energy of sub-eV is needed. This can be obtained through excitation or ionization
 of impurities
naturally existing in a high-purity Ge detector with phonons. 
The main impurities remaining in a high-purity p-type Ge crystal 
after many passes of zone refining and crystal growth 
are boron (B), aluminum (Al), gallium (Ga), and phosphorus (P). The net impurity level is dominated by 
the sum of B, Al, and  Ga in a p-type crystal. 
To make a high-purity Ge detector, the net impurity level is usually controlled within 
3$\times$10$^{10}$/cm$^{3}$, determined by Hall Effect system at 77 K. 
For aiming ionization or excitation of impurities, the detector must be a p-type or an n-type. In a
p-type detector, the p-type impurities are significantly higher than that of n-type or vice versa. If the Hall mobility
is measured at a level of $>$ 45,000 cm$^{2}$/Vs, the sum of p-type impurities (B, Al, and Ga) 
must be a factor of 10 higher than the level of phosphorus. Therefore, the total impurity
level can be approximately treated as 3$\times$10$^{10}$/cm$^{3}$. 
 
\begin{figure}
\includegraphics[width=0.48\textwidth]{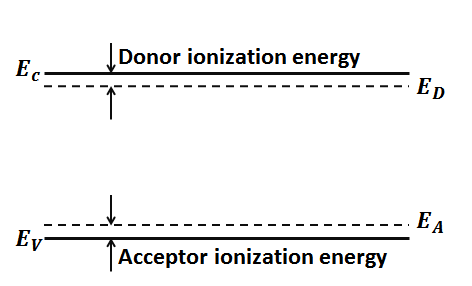}
\caption{\label{fig:bandgap}
Shown is a drawing about existing impurity changing bandgap energy. Where $E_{C}$ stands for the minimum
energy of conduction band; $E_{D}$ refers to the energy of valance of donors; $E_{A}$ is the energy of conduction 
of acceptors; and $E_{V}$ is the maximum energy of valance band. Thus, ($E_{C}$ - $E_{D}$) is the ionization 
energy of donors and ($E_{A}$ - $E_{V}$) is the ionization energy of acceptors.  
}
\end{figure}

Due to the existence of impurities, the bandgap energy can be changed
as shown in Figure~\ref{fig:bandgap}~\cite{sms,bvz}.
The ionization energies of those impurities are shown in Table~\ref{impurities}~\cite{rob}.
\begin{table}[htb]
\caption{Ionization energies of shallow impurities in Ge. }
\centering
\begin{tabular}{|c|c|c|c|c|}
\hline
 Impurity & Boron & Aluminum & Gallium & Phosphorus\\
	\hline
 	Ionization Energy & 0.0104 & 0.0102& 0.0108&0.012 \\ 
	 (eV) & & & &\\ 
	\hline
\end{tabular}
\label{impurities}
\end{table}
As can be seen from Table~\ref{impurities}, the ionization energies of impurities are all
in the range of $\sim$0.01 eV, phonons with energies of 0.037 eV and 0.026 eV can certainly ionize or excite impurities
to produce charge carriers. 
However, since the deposited energy from nuclear motion or electronic motion
is in the range of sub-eV, the ionization or excitation of impurities can only produce a few charge carriers 
per interaction induced by DM particles.
Such a small amount of charge needs to be amplified internally in order to overcome the electronic noise
in the digitization. We describe a Ge detector utilizing internal amplification of charge carriers created by the ionization of impurities below. 

\section{Development of a Detector with Internal Amplification of Charge Created by the Ionization of Impurities}
\subsection{Zone refining}
\begin{figure}
\includegraphics[width=0.48\textwidth]{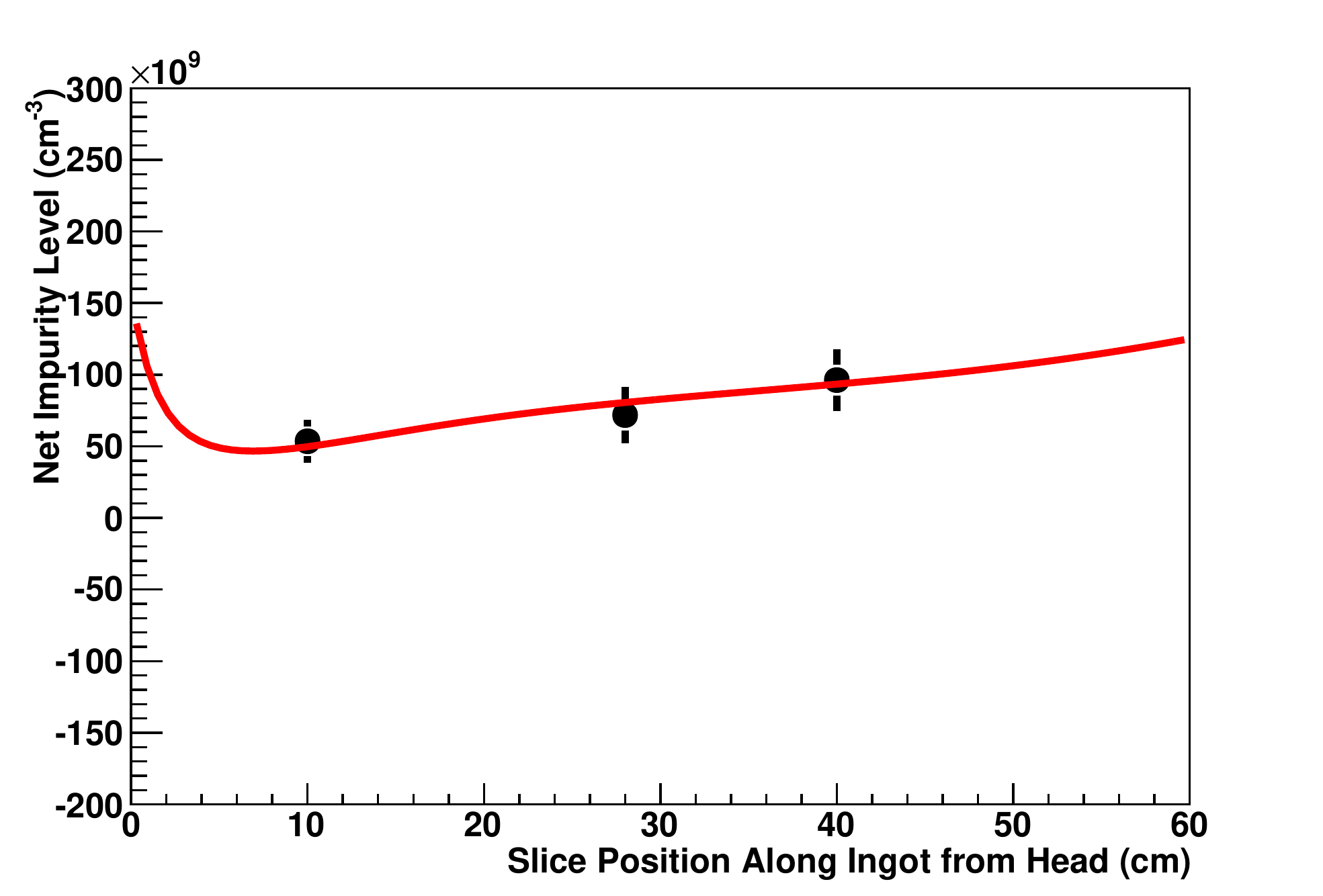}
\caption{\label{fig:zone}
Shown is a comparison between the experimental data (three dotted points in black) and theoretical prediction (equation~\ref{zone1}, solid line in red).
}
\end{figure}

Zone refining of commercially available Ge ingots is a prerequisite for growing detector-grade single crystals. We have developed zone-refining methods~\cite{gang1, gang2, gang3} that advance approaches to reducing impurity level from a commercially-achievable level of $\sim$(1-3)$\times$10$^{14}$/cm$^3$ to a level of $\sim$10$^{10}$/cm$^3$, which is needed to grow crystals without doping for novel detectors. The dominant impurities in Ge ingots are B, Al, Ga, P, and Al monoxide (AlO). The theoretical segregation coefficients between solid and liquid are 17 
for B, 0.073 for Al, 0.087 for Ga, 0.08 for P, and $\sim$1.0 for AlO~\cite{z1,z2}. The equation below shows (see Figure~\ref{fig:zone}) the net impurity level of $\sim$10$^{10}$/cm$^3$ that has been achieved after 10 passes.
\begin{equation}
\label{zone1}
|N_{A} - N_{D}| = \sum_{i} C_{i}(x)[1-(1-k_{i})exp(-\frac{k_{i}x}{L})]^n,
\end{equation}
where $|N_{A} – N_{D}|$ is the remaining net impurity in the Ge ingot after zone refining, $x$ is the length of the Ge ingot, $C_{i}$(x) is the initial purity, $i$ runs from 1 to 4 for B, Al, Ga, and P in the Ge ingot respectively, $k_{i}$ is the effective segregation coefficient in Ge, $L$ is the width of the melting zone, and $n$ is the number of passes. The value 
of $k_{i}$, depending on the zone speed, zone width, and the number of passes, is determined individually by the experimental data. The zone-refined Ge ingots usually appear to be p-type after 10 passes and can be directly used to grow a p-type Ge crystal or intentionally doped to grow an n-type crystal.  
\subsection{High-purity Ge crystal growth}
\begin{figure}
\includegraphics[width=0.48\textwidth]{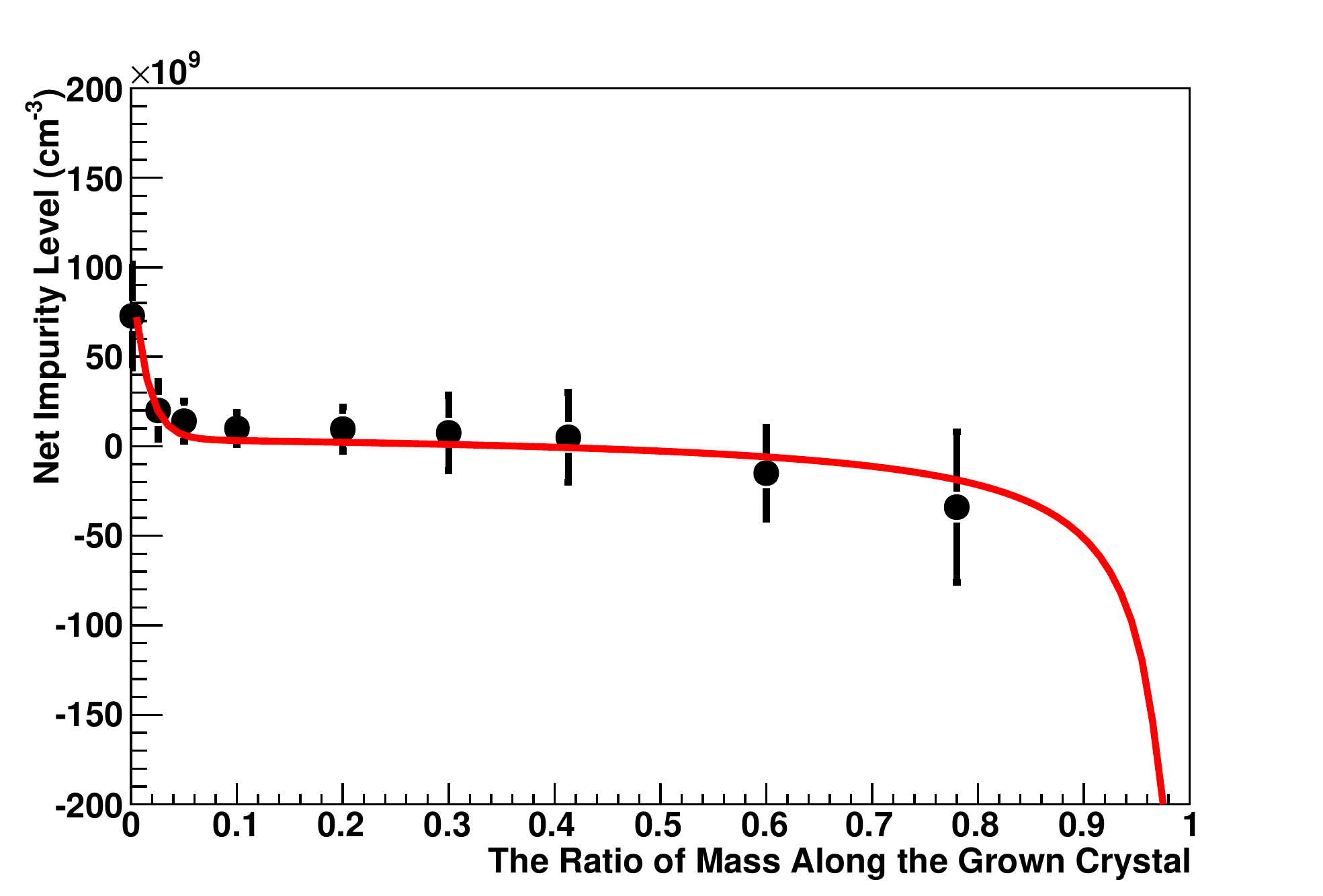}
\caption{\label{fig:crystal}
Shown is a comparison between the experimental data (dotted points in black) and theoretical prediction (equation~\ref{crystal1}, solid line in red) for a grown crystal with large portion to be detector-grade crystal.
}
\end{figure}

Large single crystals of Ge grow using the Czochralski technique~\cite{czo} at the University 
of South Dakota (USD). Since 2010, we have developed methods~\cite{usd1, usd2, usd3, usd4}
 to improve the quality of large size Ge crystals, demonstrating our
ability to control the parameters for growth of low-dislocation (3,000 – 7,000 etch pits/cm$^{2}$), large diameter
($\sim$12 cm), and high-purity Ge single crystals ($\sim$10$^{10}$/cm$^{3}$) for fabricating into detectors.
Due to the differences in the segregation coefficients of impurities, including Al, B, Ga, P, and AlO inside the melt of Ge, the level of impurities in a grown crystal exhibits a distribution along the axis of crystal. The net impurity can be calculated using the formula below: 
\begin{equation}
\label{crystal1}
|N_{A} - N_{D}| = \sum_{i} C_{i}\times k_{i}\times(1-g)^{(k_{i}-1)},
\end{equation}
with $|N_{A}-N_{D}|$ being the net impurity level after the growth, $C_{i}$ the initial impurity of B, Al, Ga, and P before the growth, $i$ representing B, Al, Ga, and P, $k_{i}$ the effective segregation coefficient for B, Al, Ga, and P respectively, and $g$ being the fraction of crystal. A comparison between the Hall effect measurements (data points in black) and the fitted function (solid line in red), using eq.~\ref{crystal1}, is shown in Figure~\ref{fig:crystal}, where a high quality crystal with diameter up to 10 cm was achieved. The measured mobility is greater than 45,000 cm$^2$/Vs along both the radial and axial directions. This indicates that the distribution of impurities inside crystal is uniform, which is critical to the proposed detector technology in this paper. 
\subsection{Internal charge amplification} 
\begin{figure}
\includegraphics[width=0.48\textwidth]{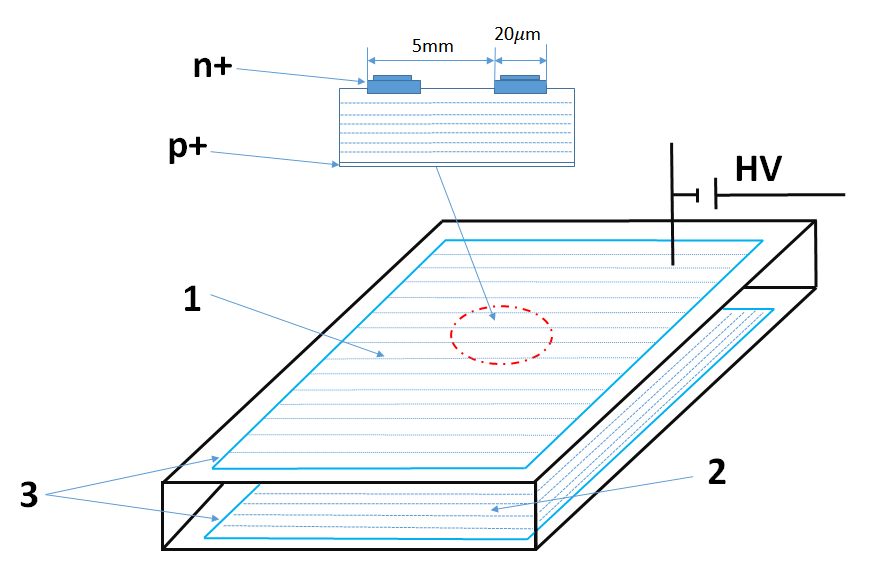}
\caption{\label{fig:planar}
Shown is a Ge detector with internal amplification~\cite{sta}. The upper
parts are: 1 – anode strips, 2 – cathode,
3 – guard electrodes, the scheme of n+
and p+ - layers.
}
\end{figure}

Using a grown crystal with mobility greater than 45,000 cm$^{2}$/Vs and a net impurity level of less than
3$\times$10$^{10}$/cm$^{3}$, a planar detector with internal amplification is designed as suggested by 
Starostin and Beda~\cite{sta}. 
In a high-purity Ge crystal with a sensitivity volume of
$\sim$190 cm$^{3}$, the critical electric field, $E_{cr}$, can be obtained at a level
of greater than 10$^{4}$ V/cm for a planar detector. 
The amplification
factor can be estimated as $K=2^{h/l}$, where $h$ is the length of
avalanche region and $l$ is the free electron path of inelastic scattering.
The value of $l$ in Ge at $\sim$ 4 K is about 0.5 $\mu$m and $h$ can be
5 $\mu$m for a planar detector of 3 cm thick, as shown in Figure~\ref{fig:planar}.
Thus, it is possible to achieve a value of $K$ = 10$^{3}$ with a threshold
as low as 0.1 eV to guarantee a low-energy threshold.

A Ge detector with internal avalanche amplification can be fabricated
at a mass of about 1.0 kg, as shown in Figure~\ref{fig:planar}. The detector can
be made of a high-purity (an impurity level of (1-3)$\times$10$^{10}$/cm$^{3}$) multistrip 
planar Ge detector and has a
dimension of 9 $\times$ 7 $\times$ 3 cm$^{3}$. There can be 15 anode strips fabricated 
using the photomask method, at a
width of 20 $\mu$m and a length of 7 cm. The expected cathode area will be 9 $\times$ 7 cm$^{2}$ and the fiducial
volume will be approximately 190 cm$^{3}$. Guard electrodes in the anode and cathode planes can also be
designed. To fabricate a Ge detector with internal amplification, one must: (1) use  a Ge
 crystal that guarantees
a uniform distribution of impurities to provide a homogeneous electric field near the anode; (2) create a wide
shallow junction layer under the strips so that the electric field near the strips is defined by junction dimensions;
and (3) guarantee reliable cooling of the crystal, since the critical electric field and amplification
factor depend on the free path of charge carriers, which in turn depend on the temperature.

\subsection{Ionization of impurities}
\subsubsection{Propagation of phonons}
\begin{figure}
\includegraphics[width=0.48\textwidth]{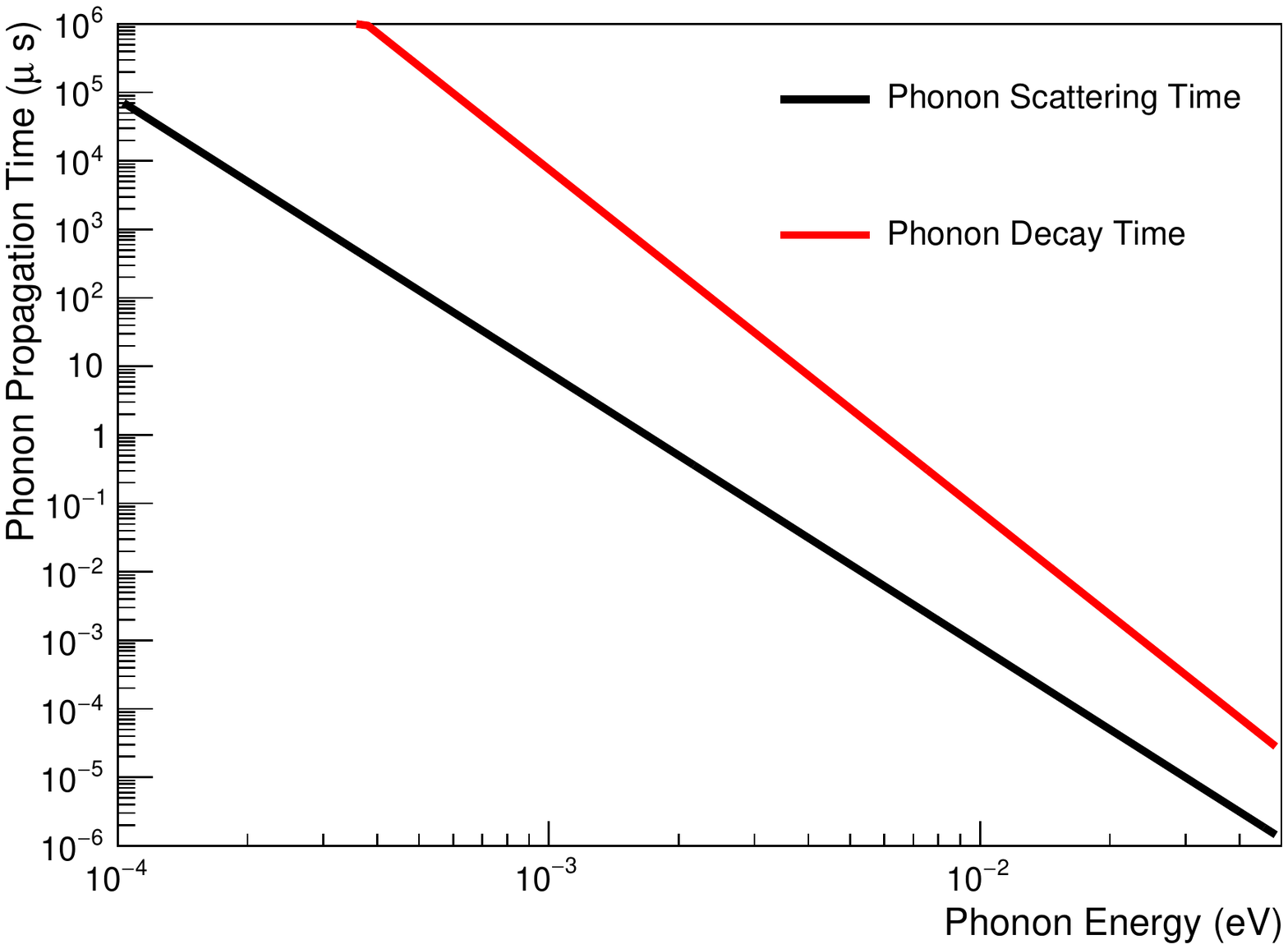}
\caption{\label{fig:propagationtime}
Shown is the timescale for the elastic scattering and the anharmonic decay process as a function of phonon energy.  
}
\end{figure}

Phonons (LA, TA, LO, TO) created directly by recoiling particles are energetic and thus subject to~\cite{prb41}: (1) elastic scattering and (2) spontaneous anharmonic decay. The former prevents energetic phonons from propagation with a scattering time, t$_s$ = $\frac{1}{A\nu^{4}}$~\cite{mem}, where A = 3.67$\times$10$^{-41}s^{3}$ is the calculated isotope-scattering constant~\cite{sta1} and $\nu$ is the frequency of phonons.  The latter down-converts phonons with a decay time, t$_{d}$ = $\frac{1}{B\nu^{5}}$, where B = 1.61$\times10^{-55}s^{4}$ is the calculated mode-averaged anharmonic-decay constant~\cite{sta1}. In the anharmonic decay process,  an original phonon splits into two lower-energy phonons (roughly equal frequency) and thus are able to propagate quasi-ballistically inside crystal~\cite{jap}. The timescale for the elastic scattering and the anharmonic decay process is shown in Figure~\ref{fig:propagationtime}.

\begin{figure}
\includegraphics[width=0.48\textwidth]{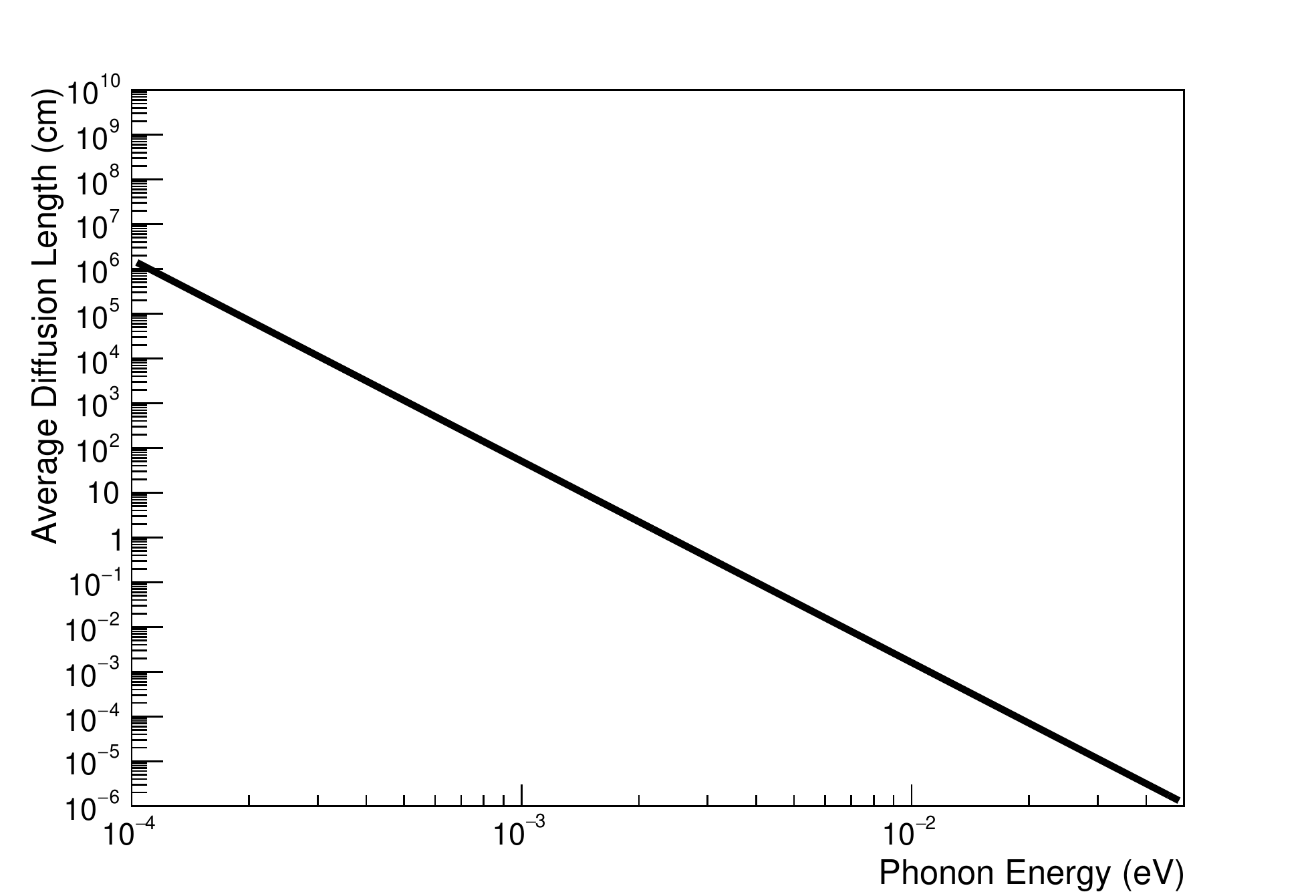}
\caption{\label{fig:propagation}
Shown is the average distance diffused before an anharmonic decay process as a function of phonon energy.  
}
\end{figure}

As can be seen in Figure~\ref{fig:propagationtime}, for the phonons with energies greater than 0.01 eV, the timescale of the elastic scattering is in the order of picoseconds. However, the anharmonic decay time is in the order of microseconds. Correspondingly, the average distance diffused before an anharmonic decay~\cite{mem} can be expressed as:
\begin{equation}
\label{diffused} 
d = \sqrt{(\frac{v_s^2}{3}t_{s}(\nu_{parent})t_d(\nu_{parent}))}, 
\end{equation}
where $\nu_{parent}$ is the frequency of any parent phonon before the anharmonic decay process and d is in the oder of $\mu$m for high energy phonons and $\sim$cm for lower energy phonons, respectively, as shown in Figure~\ref{fig:propagation}.  

\subsubsection{Phonons interactions with impurities}
The detector can be fully depleted with a reversed bias of 4000 volts at 77 K. After the depletion, the
detector can be cooled down further to $\sim$1.5$-$4 K, which freezes all impurities into neutral impurities. At this
temperature regime ($\sim$1.5$-$4 K), the scattering mechanism is dominated by neutral impurities~\cite{neu}.
The phonons will mainly scatter off neutral impurities to excite or ionize 
B, Al, Ga, and P and generate charge carriers through the following reactions:
\begin{equation}
D^{X} + \bigtriangleup E_{D} \longrightarrow e^{-} + D^{+},
\end{equation}
and 
\begin{equation}
A^{X} + \bigtriangleup E_{A} \longrightarrow h^{+} + A^{-}. 
\end{equation}
$D^{X}$ represents neutral donors; $\bigtriangleup E_{D}$ stands for the energy absorbed by neutral donors;
the produced charge carrier, $e^{-}$, is to the conduction band described in Figure~\ref{fig:bandgap}.
Similarly, $A^{X}$ denotes neutral acceptors; $\bigtriangleup E_{A}$ stands for the energy absorbed by neutral acceptors;
the produced charge carrier, $h^{+}$, is to the valance band described in Figure~\ref{fig:bandgap}. 
Then, the charge carriers will be drifted toward to the anode strips, where they will be 
accelerated to generate an avalanche effect and hence amplify the charge by a factor of $\sim$1000.

\section{Evaluation of the Detector Sensitivity}
\subsection{Signal from the ionization or excitation of impurities}
Ionization or excitation of impurities in Ge has been observed in many experiments~\cite{aph, nsc, dls, wpi}. The ionization or excitation cross section of both over-charged $D^{-}/A^{+}$ and neutral $D^{0}/A^{0}$ impurities was measured to be $<\sigma>$=5$\times$10$^{-13}$ cm$^{2}$ by Phipps et al.~\cite{aph}. We can estimate the absorption probability for a phonon (LA, TA, LO, and TO) propagating in a given Ge detector discussed above:
\begin{equation}
\label{exciationrate}
P = 1 - exp(-d/\lambda),
\end{equation}
where $P$ is the absorption probability, d is the average distance diffused before a subsequent anharmonic decay process (eq.~\ref{diffused}), and $\lambda$ = $\frac{1}{<\sigma>\times N_{A}}$ is the mean free path of phonons with $N_{A}$ the net impurity level in a given detector. 

\begin{figure}
\includegraphics[width=0.48\textwidth]{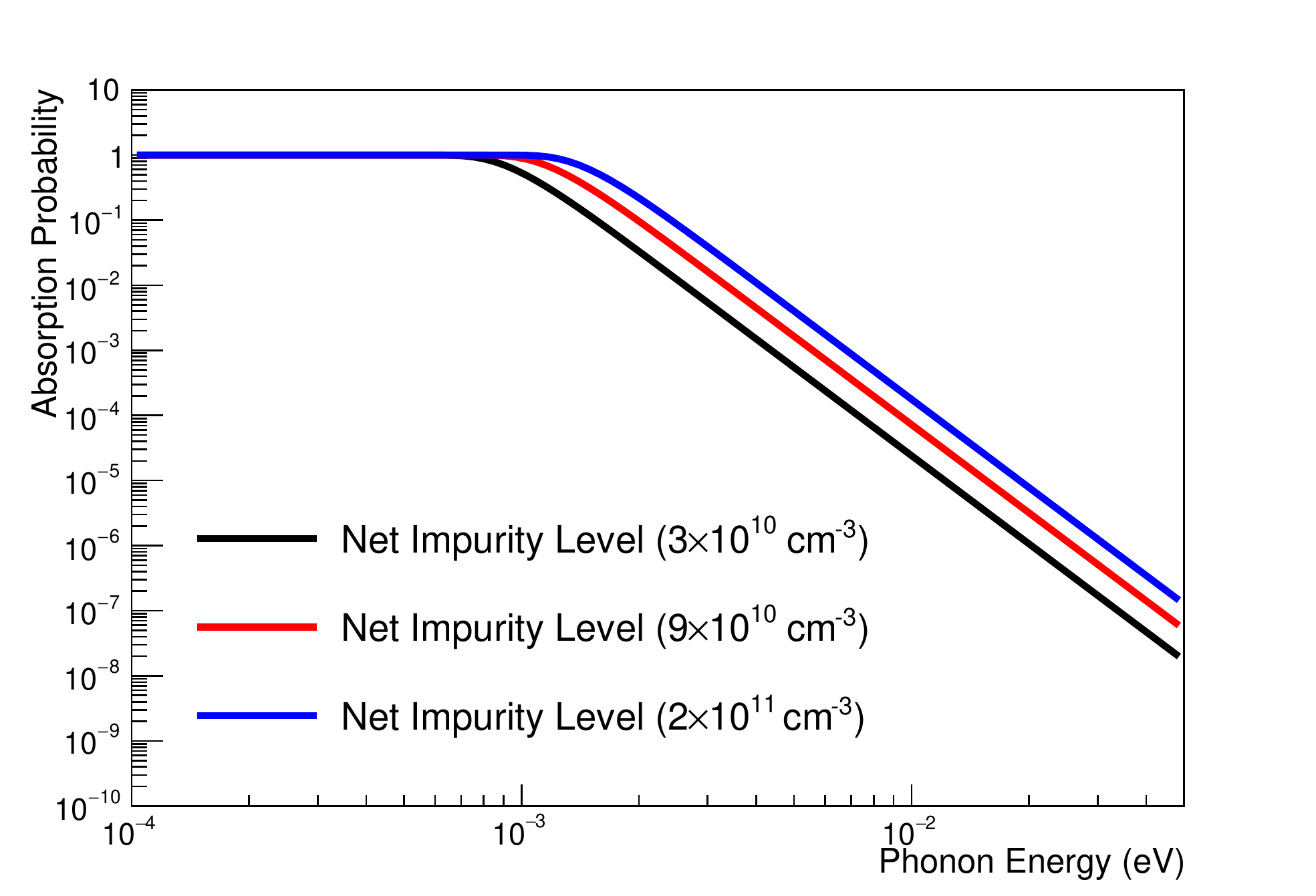}
\caption{\label{fig:probability}
Shown is the absorption probability as a function of phonon energy in a given Ge detector. Three impurity levels are displayed and the difference in the absorption probability is seen when the energy of phonons is greater than 0.001 eV.   
}
\end{figure}

\begin{figure}
\includegraphics[width=0.48\textwidth]{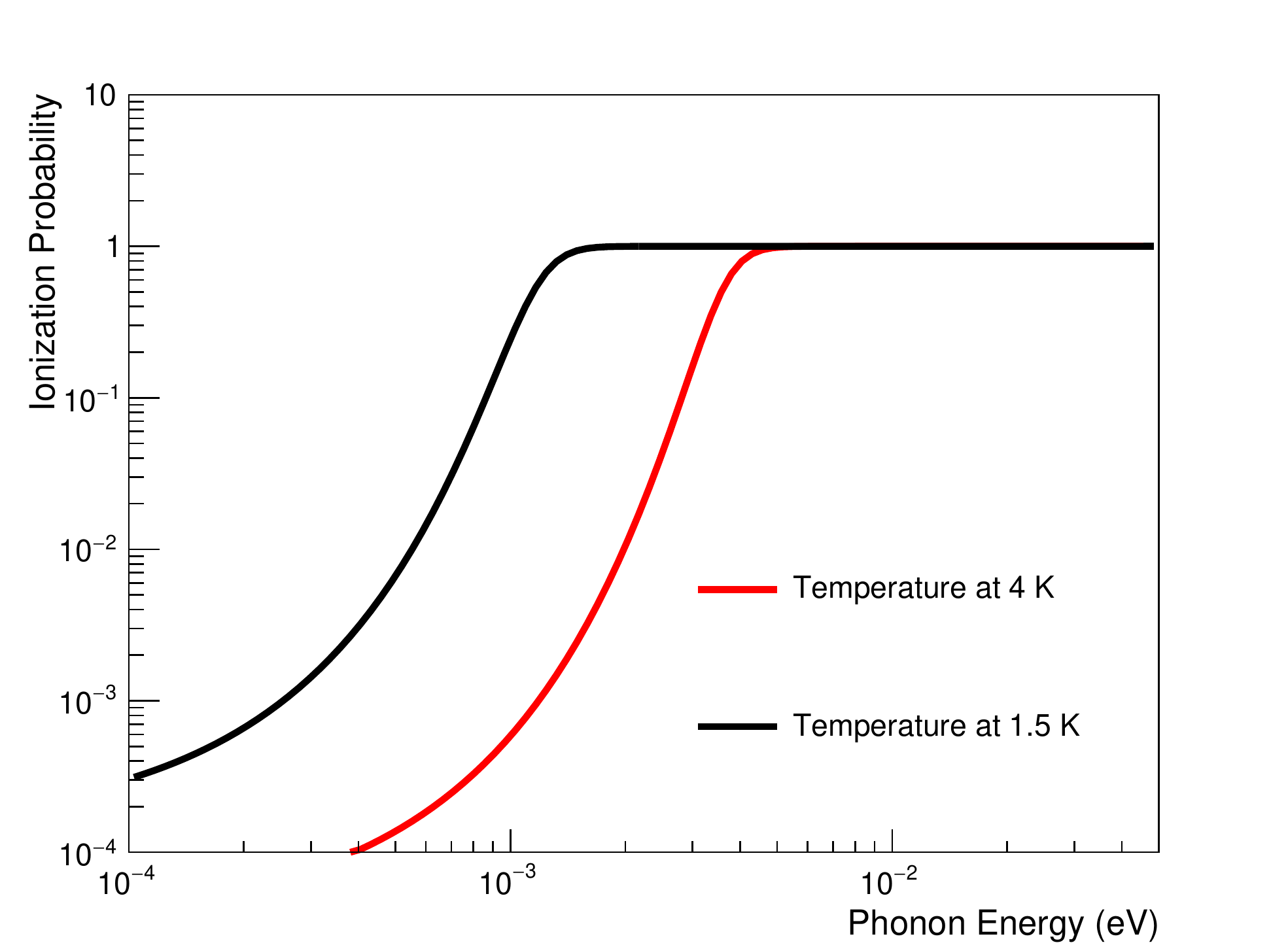}
\caption{\label{fig:ionization}
Shown is the ionization or excitation probability as a function of phonon energy in a given Ge detector.    
}
\end{figure}

Figure~\ref{fig:probability} shows the absorption probability for three impurity levels, 3$\times$10$^{10}$/cm$^3$, 9$\times$10$^{10}$/cm$^3$, and 2$\times$10$^{11}$/cm$^3$, which can be used to obtain a desirable efficiency for ionization or excitation of phonons. 
Once a phonon with energy, E$_{D}$ or E$_{A}$, is absorbed by a D$^{X}$ or A$^{X}$ state, the ionization or excitation probability can be estimated using Fermi-Dirac statistics~\cite{fermi}. Since the designated detector in this work is a p-type, the probability for a neutral acceptor state to be ionized is described below:
\begin{equation}
\label{donor}
f(E_{A}) = 1 - \frac{1}{1+4e^{(E_{A}-E_{F})/k_{B}T}},
\end{equation}
where E$_{F}$ is Fermi energy level and (E$_F$ - E$_{V}$) = k$_{B}$Tln($\frac{N_{V}}{N_A}$) with N$_{V}$ = 2(2$\pi$m$^{*}$k$_{B}$T/h$^2$)$^{3/2}$ being the effective states, m$^{*}$ is the effective mass of a hole, k$_{B}$ is the Boltzmann constant. Figure~\ref{fig:ionization} shows the ionization or excitation probability of a neutral acceptor state when a phonon is absorbed. Note that this probability depends strongly on the temperature. Therefore, one can vary temperature from 1.5 K to 4 K to achieve a desirable ionization or excitation probability at the detection threshold. As can be seen in Figure~\ref{fig:ionization}, the energies of phonons that are smaller than 0.01 eV, the critical ionization energy of impurities, can have some level of probability to ionize or excite acceptor states due to the acoustic phonon assisted tunneling~\cite{dargys}. 

If one assumes a detection threshold of 0.1 eV, which corresponds to $\sim$4 phonons with  an average energy of 0.026 eV per phonon, we can estimate the number of charge carriers created by the ionization or excitation of impurities using the following equation:
\begin{equation}
\label{detection}
N_{carriers} = \sum_{i} n_{i}P_{i}f_{i},
\end{equation}
where n$_{i}$ is the number of the i$^{th}$ phonons, P$_{i}$ is the absorption probability (eq.\ref{exciationrate}) for the i$^{th}$ phonons, and f$_{i}$ is the ionization or excitation probability (eq.~\ref{donor}) for the i$^{th}$ phonons. Note that the i$^{th}$ phonons refer to the i$^{th}$ generation phonons after the i$^{th}$ anharmonic decay, which splits a parent phonon into two phonons with almost the same energy. 

\begin{figure}
\includegraphics[width=0.48\textwidth]{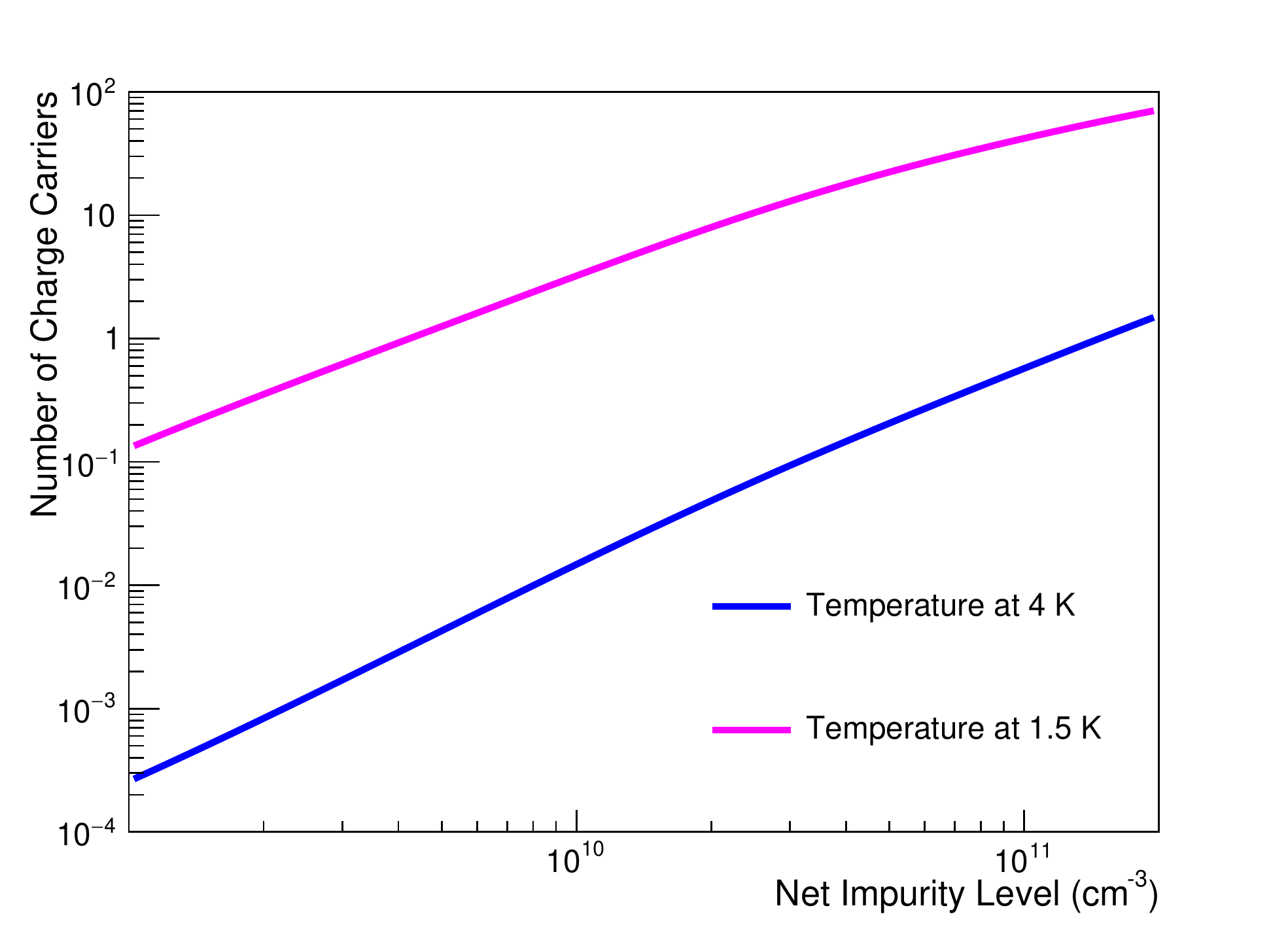}
\caption{\label{fig:carriers}
Shown is the number of charge carriers created by the ionization or excitation of impurities as a function of impurity level in a given Ge detector. Two temperatures, 1.5 K and 4 K, are displayed. 
}
\end{figure}

Figure~\ref{fig:carriers} shows the number of charge carriers created by the ionization or excitation of phonons with a total energy of 0.1 eV from recoiling particles. At the impurity level of 3$\times$10$^{10}$/cm$^{3}$, more than 12 charge carriers can be generated by 4 parent phonons with an initial average energy of 0.026 eV per phonon after several generations of the elastic scattering and the anharmonic decay process,  when the detector is operated at 1.5 K. With an impurity level of 2$\times$10$^{11}$/cm$^{3}$, at least one charge carrier can be produced when the detector is operated at 4 K. 
Therefore,  the phonons created by recoiling particles can be detected at a level of $>$50\%, if one takes into account the loss of phonons due to the internal reflection at the boundary with the specific Ge detector discussed above. 

It is worth mentioning that the phonons can be generated during the avalanche process in which the accelerated charge carriers have sufficient kinetic energies (greater than the Ge bandgap) to produce e-h pairs. Since Ge is an indirect bandgap, the generation of an e-h pair requires momentum conversation~\cite{wei}. Consequently, there are phonons accompanying the generation of e-h pairs. The average phonon energy is estimated to be about 0.00414 eV~\cite{wei}, which has small probability to ionize or excite impurities, according to eq.~\ref{detection}. Since there are as many as a few hundreds of phonons produced per e-h pair, the charge carriers created by these phonons could cause electric breakdown during the avalanche process. However, the avalanche process is within 100 ns and the ionization or excitation of impurities by those phonons are delayed by about 10 $\mu$s, as can be seen from Figure~\ref{fig:propagationtime}. This allows us to decrease the E-field for stopping avalanche for about 1 ms right after the primary pulse generated by the avalanche process. Therefore, electric breakdown can be avoided by a strategic operation. 

\subsection{Evaluation of backgrounds}
In the detection region of interest (sub-eV to 100 eV),  the expected background rates  are essentially negligible from natural radioactivities and muon-induced processes due to  the ability of detecting a single  charge carrier. This can be  estimated with an achieved background rate from a Ge DM experiment, such as CDEX~\cite{cdex}. CDEX reported a background rate of (4.09$\pm$1.71) kg$^{-1}keV^{-1}d^{-1}$, which can be translated to be (4.09$\pm$1.71)$\times$10$^{-3}$ kg$^{-1}eV^{-1}d^{-1}$. This rate is dominated by cosmogenic backgrounds such as tritium and $^{68}$Ge, as shown by the M{\sc{ajorana}} D{\sc{emonstrator}}~\cite{mjd}. This rate can be further reduced to be about 0.04 kg$^{-1}eV^{-1}y^{-1}$. Neutrino elastic scattering is another source of backgrounds and is estimated in the level of $\sim$0.001 kg$^{-1}eV^{-1}y^{-1}$\cite{res, jbi}.

The bulk thermal noise depends on the thermal energy, which is about 0.00033 eV at 4 K. The excitation probability with such a small thermal energy is estimated to be at a level of $\sim$10$^{-4}$. This is completely negligible. Another possible source of background is the single electron injection from electrodes. The materials with higher values of work-function will be chosen to minimize this source of background.  It is worth mentioning that breakdown observed at fields of order of $\sim$30 V/cm in the SuperCDMS-type detector at a base temperature of $\sim$30 mK was not due to 
the impact ionization of the bulk of the crystals~\cite{luke}. This also indicates that thermal noise will not be a significant source of background.  

\subsection{Projected sensitivity}
We evaluate the sensitivity of the specific detector, discussed above, to MeV-scale DM-nucleus scattering using the most common way,
which displays direct detection results based on a differential rate with spin-independent 
and isospin-conserving interactions:
\begin{equation}
\frac{dR}{dE}(E,t) = \frac{\rho_{0}}{2\mu_{\chi N}^2\cdot m_{\chi}}\cdot\sigma_{0}\cdot A^2\cdot F^2\int_{v_{min}}^{v_{esc}}\frac{f(\vec{v},t)}{v}d^{3}v,
\label{eq1}
\end{equation}
where $\rho_{0}$ = 0.4 GeV/cm$^{3}$ (local DM density); $\mu_{\chi N}$ is the DM-nucleus reduced mass; $m_{\chi}$ is the mass of nucleus; $\sigma_{0}$ is the DM-nucleus cross section at the zero momentum transfer; $A$ is the number of nucleons per nucleus; $F$ is the nuclear form factor; $v_{esc}$ = 544 km/s (the escape velocity); and the minimum velocity is defined as:
\begin{equation}
v_{min} = \sqrt{\frac{m_{\chi}E_{th}}{2\mu_{\chi N}^2}},
\label{eq2}
\end{equation}
with $E_{th}$ being the detection energy threshold. Finally, the DM velocity profile is commonly described by an isotropic Maxwell- Boltzmann distribution:
\begin{equation}
\label{eq3}
f(\vec{v},t) = \frac{1}{\sqrt{2\pi \sigma}}exp(-\frac{\vec{v}^2}{2\sigma^2}),
\end{equation}
where $\sigma$ = $\sqrt{3/2}v_{c}$ and $v_{c}$ = 220 km/s. For spin independent interactions, the cross-section at zero momentum transfer can be expressed as:
\begin{equation}
\label{spin}
\sigma_{0}^{SI}=\sigma_{p}\cdot\frac{\mu_{\chi N}^{2}}{\mu_{p}^{2}}\cdot[Z\cdot f^{p} + (A-Z) \cdot f^{n}]^2,
\end{equation}
where $\sigma_{p}$ is the DM-nucleon cross section; $f^{p}$ and $f^{n}$ are the contributions of protons and neutrons to the total coupling strength, respectively; $\mu_{p}$ is the WIMP-nucleon reduced mass; and $A$ is the nuclear mass number. It is common that, $f^{p}$ = $f^{n}$ is assumed and the dependence of the cross-section with the number of nucleons $A$ takes an A$^{2}$ form. For light DM, the momentum transfer is small and the nuclear form factor, $F$, can be assumed to be 1. 
The total event rate is given by:
\begin{equation}
\label{total}
R = \frac{\rho_{0}}{m_{\chi}}(2/\pi)v_{\chi}\sigma_{0}^{SI}(6\times10^{26}/A) s^{-1}kg^{-1},
\end{equation}
where $v_{\chi}$ is the velocity of DM. 

In the case of DM-electron scattering under the light mediator framework, the event rate and sensitivity are discussed in detail by Essig et al~\cite{res,ess1} for different targets including Ge, which shows the best sensitivity. It is expected that the sensitivity can be improved with the detector technology described in this paper, since the detection threshold is 0.1 eV, which is a factor of 30 lower than 3 eV used in the work of Essig et al.~\cite{res, ess1}. 
We summarize the theory for the DM-electron scattering mediated through a dark photon, with m$_{A'}$ = 3m$_{\chi}$,  at which we define the DM-electron scattering cross section $\bar{\sigma_{e}}$ and the DM form factor F$_{DM}$(q) as~\cite{res,ess1}:
\begin{equation}
\label{dme1}
\bar{\sigma_{e}}= \frac{16\pi\mu_{\chi e}^{2}\alpha\epsilon^2\alpha_{D}}{(m_{A'}^2+\alpha^2m_{e}^2)^2}\simeq\{\begin{array}{ll} \frac{16\pi\mu_{\chi e}^2\alpha\epsilon^2\alpha_{D}}{m_{A'}^4}, & m_{A'}>>\alpha m_{e}, \\
 \frac{16\pi\mu_{\chi e}^2\alpha\epsilon^2\alpha_{D}}{(\alpha m_{e})^4}, & m_{A'}<<\alpha m_{e}.
 \end{array}
\end{equation}
\begin{equation}
F_{DM}(q) = 
\frac{m_{A'}^2+\alpha^2 m_{e}^2}
{m_{A'}^2+q^2}
\simeq\{\begin{array}{ll} 
1, & m_{A'}>>\alpha m_{e},\\
\frac{\alpha^2m_{e}^2}{q^2}, & m_{A'}<<\alpha m_{e}.
\end{array}
\end{equation}
Where $\alpha_{D}\equiv g_{D}^2/4\pi$ with g$_{D}$ the Abelian gauge coupling constant, $\epsilon$ is a small coupling constant to ordinary charged particles through kinetic mixing with the photon (m$_{A'}$)~\cite{bho,pga}, $\mu_{\chi e}$ is the DM-electron reduced mass, and q is the momentum transfer
between the DM and electron.

We show projections for DM-nucleus and DM-electron scattering through a dark photon mediator. The DM-electron projections have been converted to DM-nucleus projections using:
\begin{equation}
\label{dmedmn}
\bar{\sigma_{e}}=4\frac{\mu_{\chi e}^2}{\mu_{\chi N}^2} \sigma_{p}.
\end{equation}

Note that, under the heavy mediator limit, the DM-electron and the DM-nucleon scattering cross section are related through:
\begin{equation}
\label{related}
\frac{\sigma_{p}}{\sigma_{e}} \simeq (\frac{\mu_{p}}{\mu_{e}})^2,
\end{equation}
where $\mu_{p}$ and $\mu_{e}$ represent the DM-nucleon reduced mass and DM-electron reduced mass, respectively. For DM mass of $\sim$100 MeV, eq.~\ref{related} can be reduced to:
\begin{equation}
\label{related1}
\frac{\sigma_{p}}{\sigma_{e}} \sim (\frac{m_{p}}{m_{e}})^2,
\end{equation}
where $m_{p}$ is the mass of a nucleon and $m_{e}$ is the mass of an electron. As can be seen from eq.~\ref{related} and eq.~\ref{related1}, the interaction strength of the DM-electron scattering is about 4 to 6 orders of magnitude smaller than the DM-nucleon scattering. Therefore, we only present the event rate and sensitivity for the DM-nucleon scattering process, under the heavy mediator limit, in this work. 

\begin{figure}
\includegraphics[width=0.48\textwidth]{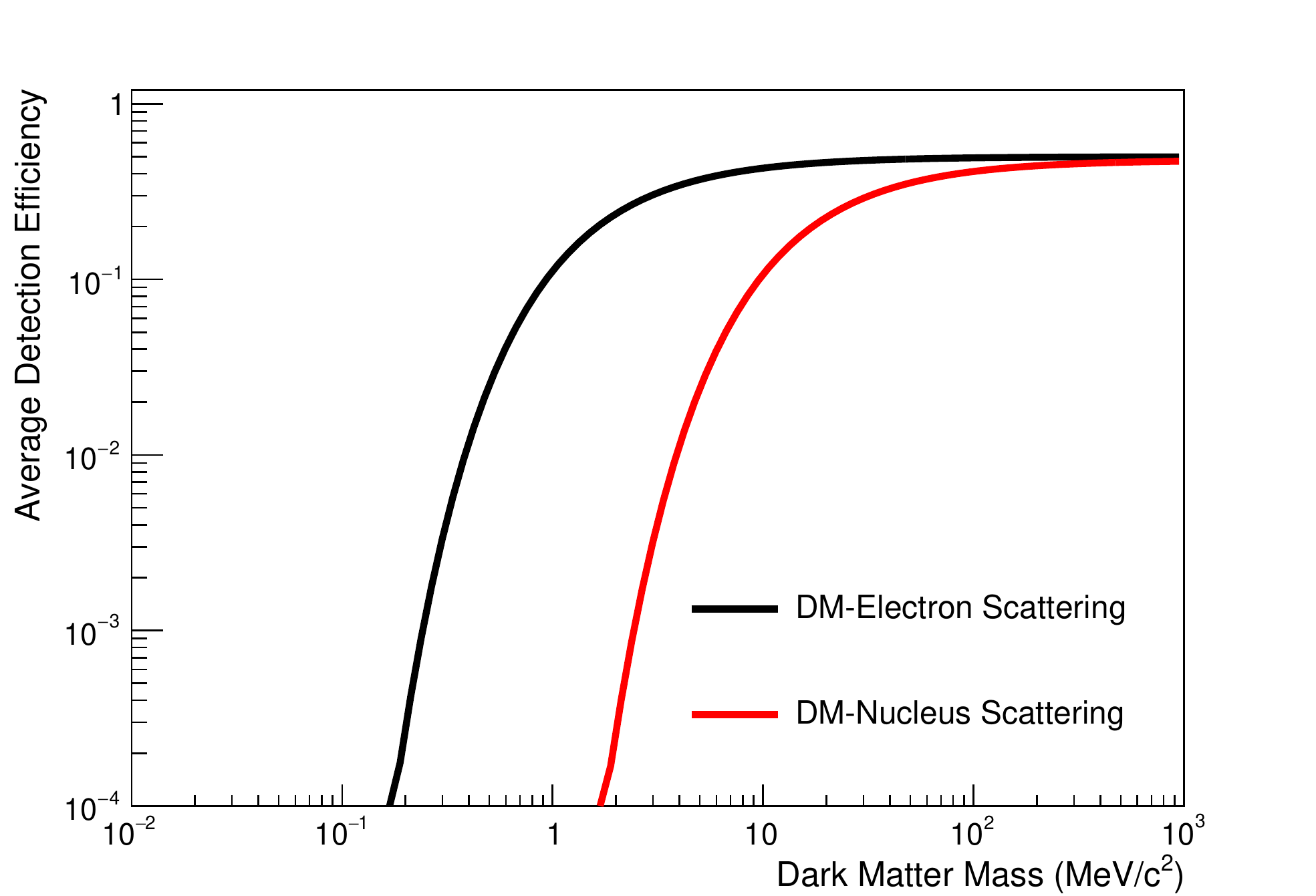}
\caption{\label{fig:eff}
Shown is the average detection efficiency, normalized to a phonon detection efficiency of 50\%, for DM-nucleus and DM-electron interactions, as a function of the mass of DM.
}
\end{figure}

In the detection of DM in the mass range of MeV-scale, the average detection efficiency as a function of DM mass is simulated for DM-nucleus and DM-electron interactions and is normalized to a phonon (0.026 eV) detection efficiency of 50\%, as shown in Figure~\ref{fig:eff}. 
We evaluate the total event rate while taking into account the average detection efficiency as a function of DM mass.  The average detection efficiency of DM-nucleus scattering decreases quickly as the mass of DM decreases. 
It does not increase to 100\% until the mass of DM approaches 1 GeV/c$^2$. However, the average detection efficiency for DM-electron interaction increases quickly as the mass of DM increases. It approaches to almost 100\% when the mass of DM is in the range of 10 MeV.
This is because the energy deposition varies with the variation of scattering angles and DM velocities for a given mass of DM particle.  In general,  the energy deposition from DM-nucleus scattering  is much less than 0.1 eV. However, when the mass of DM is  more than 500 MeV/c$^2$, the induced nuclear recoil energy can be a few hundreds of eV, as shown in Figure~\ref{fig:fullSpe}. This increases the detection efficiency of such an interaction significantly. On the other hand, the energy deposition from DM-electron is greater than 0.1 eV when the mass of DM is more than a few MeV/c$^{2}$ allowing a detection efficiency of close to 100\%. 

\begin{figure}
\includegraphics[width=0.48\textwidth]{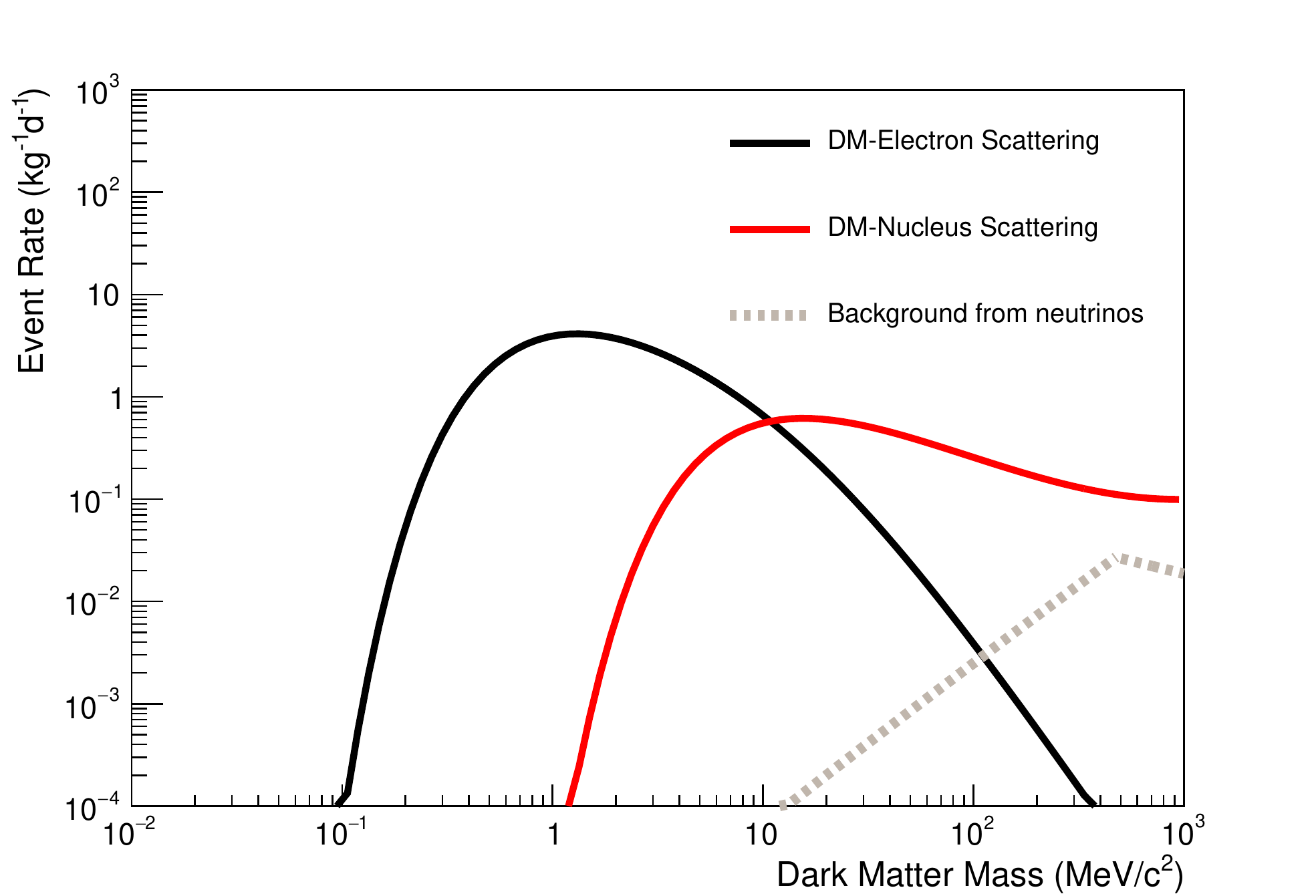}
\caption{\label{fig:totEvt}
Shown is the projected total event rate, for both DM-nucleus and DM-electron scatterings, assumed a DM-nucleon scattering of 10$^{-42}$ cm$^2$ and $\bar{\sigma_{e}}$ calculated with eq.~\ref{dmedmn}. DM-electron scattering has more events when the mass of DM is below 1 MeV/c$^2$ while comparing to DM-nucleus scattering at which more events appear when the mass of DM is above 20 MeV/c$^2$. 
}
\end{figure}

\begin{figure}
\includegraphics[width=0.48\textwidth]{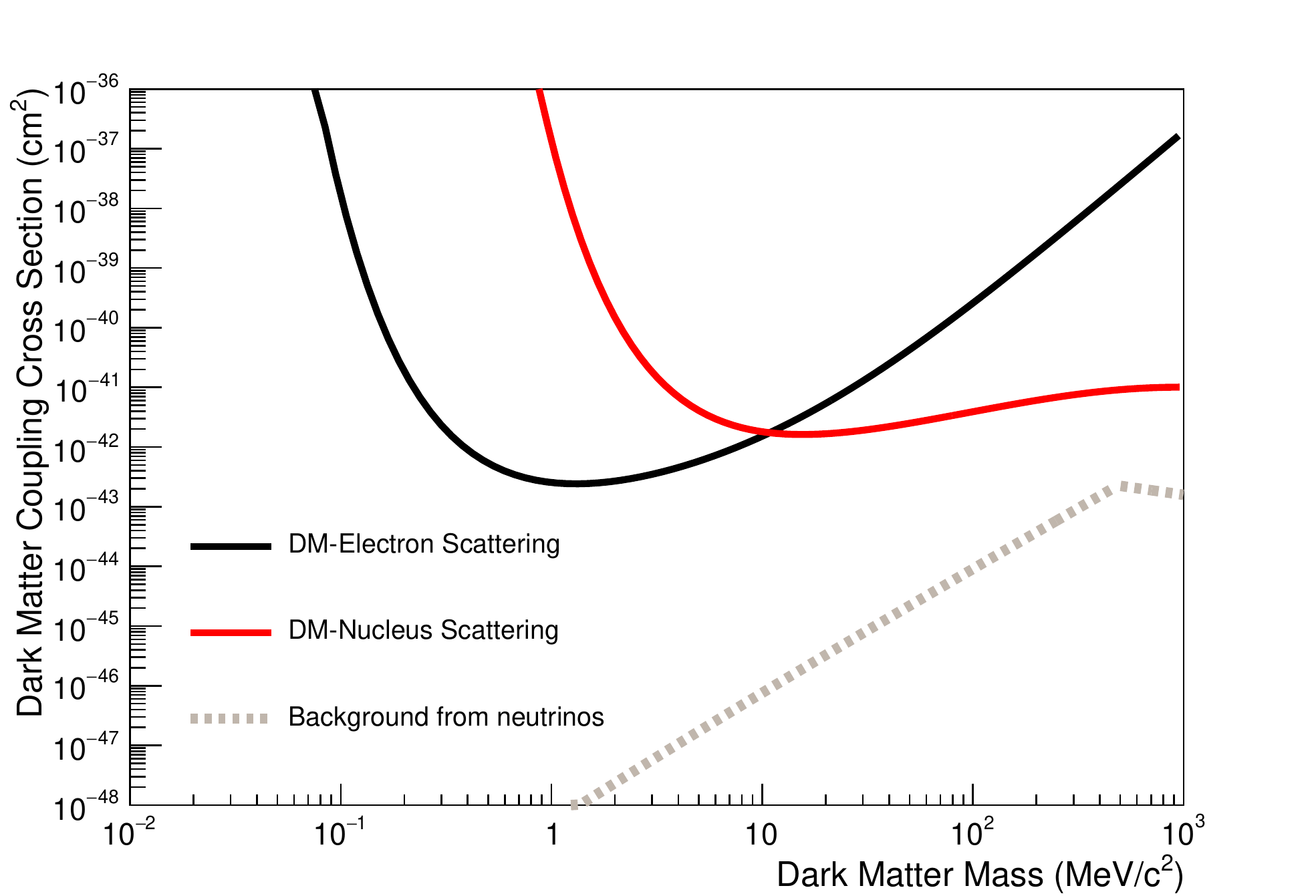}
\caption{\label{fig:sens}
Shown is the projected experimental sensitivity for 1 kg-day exposure. DM-electron scattering, under the light mediator with a dark photon, is more sensitive to the mass of DM below 1 MeV/c$^2$ while comparing to DM-nucleus scattering. 
}
\end{figure}

\begin{figure}

\includegraphics[width=0.48\textwidth]{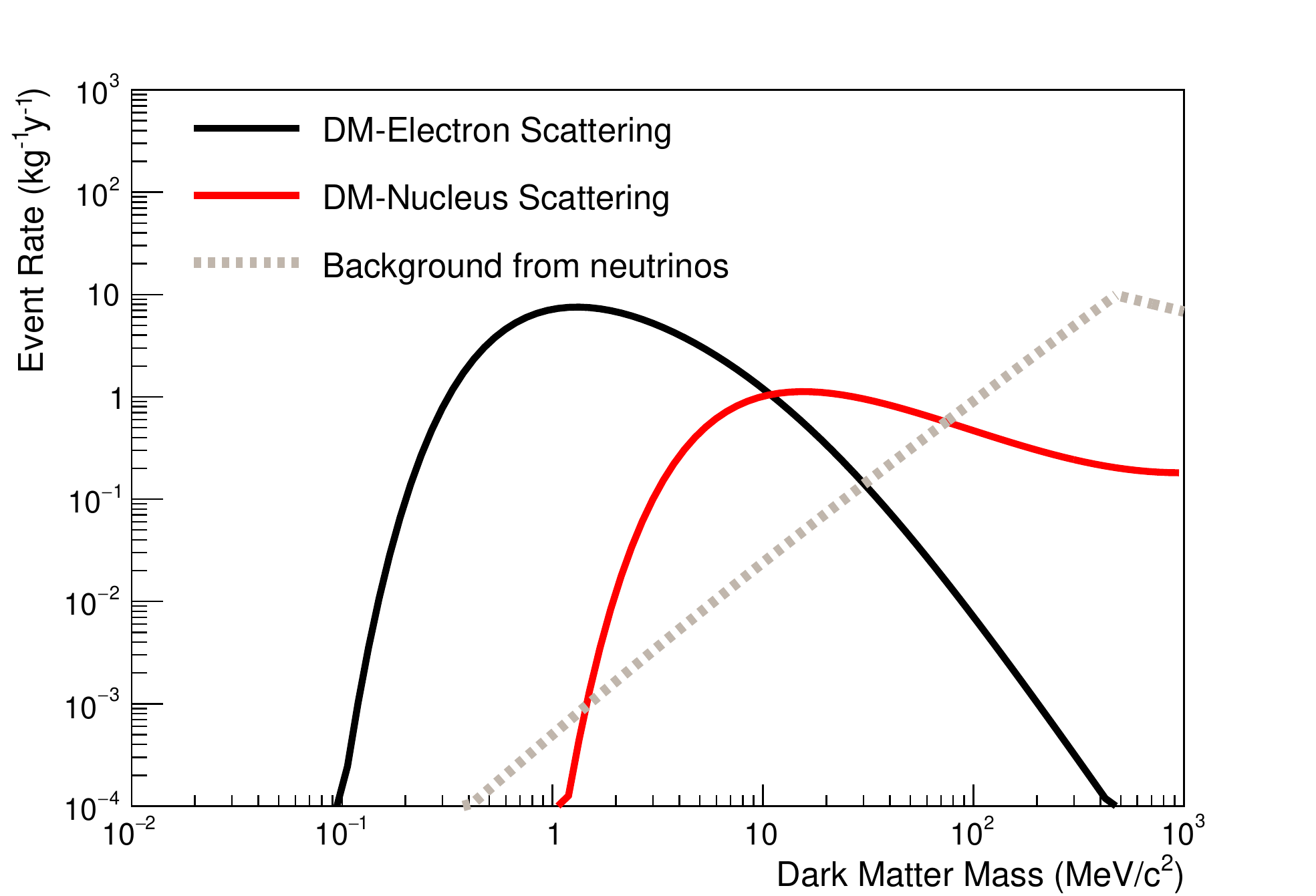}
\caption{\label{fig:back}
Shown is the projected event rate for a year exposure assumed a DM-nucleon cross section of 5$\times$10$^{-45}$ cm$^{2}$ and a DM-electron cross section calculated using eq.~\ref{dmedmn}. The background events are assumed to be constrained by neutrino induced nuclear recoils per kg per year~\cite{jbi}. 
}
\end{figure}

Figure~\ref{fig:totEvt} demonstrates the projected total event rate for 1 kg-day exposure, assumed a DM-nucleon cross-section of 10$^{-42}$ cm$^{2}$ and a DM-electron cross-section calculated using eq.~\ref{dmedmn}. 
As can be seen in Figure~\ref{fig:totEvt}, there will be a few events per day for detecting DM-electron scattering when the mass of DM is greater than 0.1 MeV/c$^2$ and a few events per day for detecting DM-nucleus scattering when the mass of DM is a few MeV/c$^2$ in a detector with only 1 kg of mass,  if we assume the DM-nucleus cross section is 10$^{-42}$ cm$^{2}$.  If we don't observe any events in the region of interest, the sensitivity for such a detector is shown in Figure~\ref{fig:sens}.
This is indeed a very sensitive experiment even with 1 kg of mass due to a much larger DM flux in the sub-MeV mass region comparing to that of the GeV mass region. 
Since the signals are single electrons in the energy region of sub-eV, the background events are negligible. For 1 kg-year exposure, as shown in Figure~\ref{fig:back}, the event rates are calculated with a DM-nucleon cross section of 5$\times$10$^{-45}$ cm$^{2}$ and a DM-electron cross section calculated using eq.~\ref{dmedmn}. 

\begin{figure}
\includegraphics[width=0.48\textwidth]{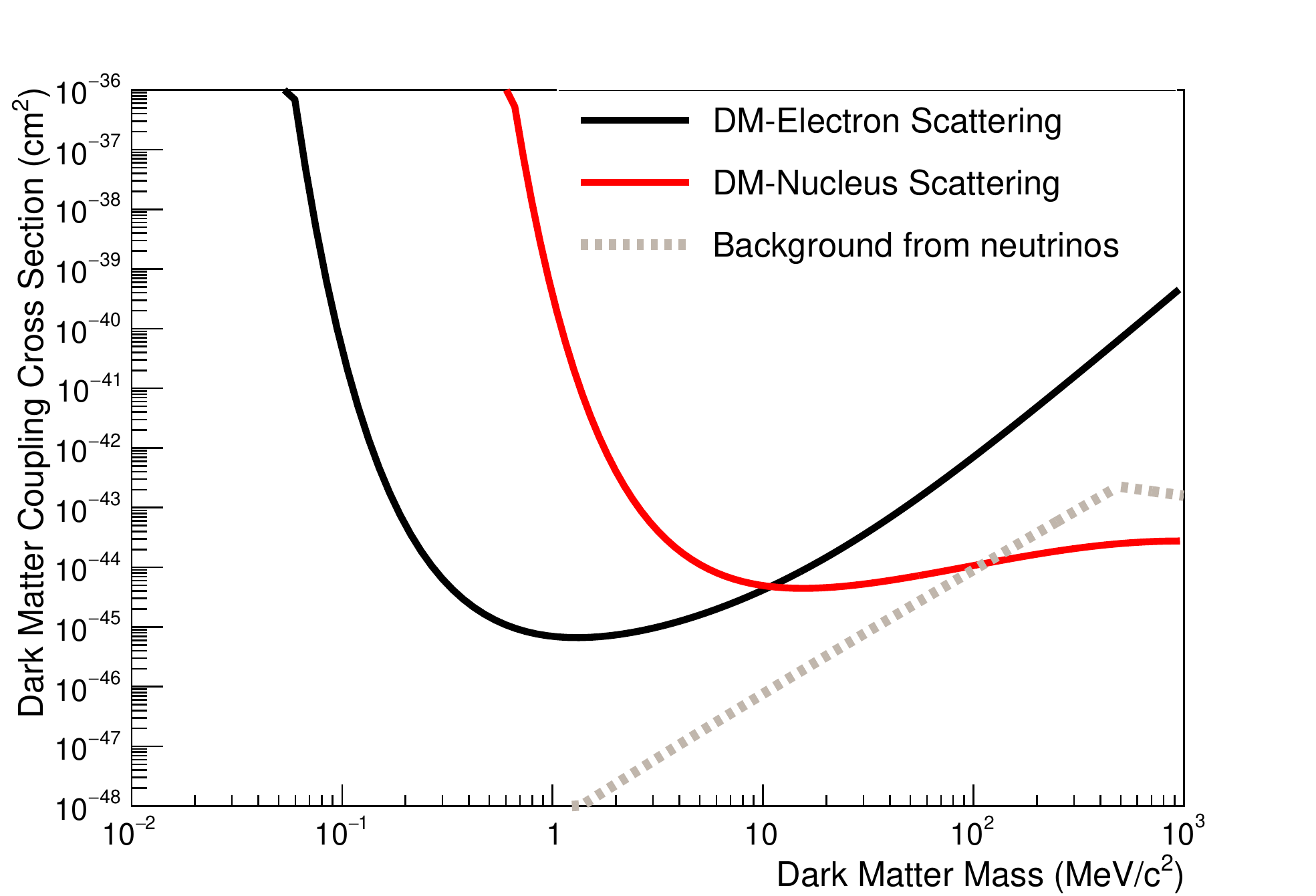}
\caption{\label{fig:sens1}
Shown is the projected experimental sensitivity for a year. The background events are assumed to be only constrained by neutrino induced background events~\cite{jbi}. 
}
\end{figure}

Figure~\ref{fig:sens1} shows the projected experimental sensitivity for one year exposure with a constrain of background events induced by neutrinos only. It demonstrates that one year exposure would set a sensitivity of DM-nucleon cross-section at a level of $\sim$5$\times$10$^{-45}$ cm$^{2}$ with a DM mass of $\sim$10 MeV/c$^2$ and a sensitivity of DM-electron cross-section at a level of $\sim$5$\times$10$^{-46}$ cm$^2$ with a DM mass of $\sim$1 MeV/c$^2$ under the light mediator limit. 
Note that, as shown in Figure~\ref{fig:back}, when the mass of DM is below 1 MeV/c$^2$, the background events from neutrinos will not constrain the sensitivity for DM-electron scattering and a 100-kg detector would be able to achieve $\sim$10$^{-48}$ cm$^2$ in a year. This 100-kg detector has discovery potential in the search for light DM with MeV-scale mass in a year.  

\section{Conclusions}
We have demonstrated a design for a very sensitive detector in the search for light DM of MeV-scale in mass. The key features of this detector are: (1) high-purity Ge detector with a net impurity level of $\sim$3$\times$10$^{10}$ cm$^{-3}$ uniformly distributed across the entire detector; (2) internal charge amplification with a gain factor of 1000; and (3) depleted detector at 77 K and continuously cooled down to $\sim$4 K. Such a detector allows phonons, created by DM through elastic scattering off either a nucleus or electrons, to excite impurities at 0.01 eV and hence create charge carriers. Taking into account the internal amplification, the best option of such a detector is n-type, which will directly generate electrons as the charge carriers when phosphorus atoms are excited by phonons. Those electrons will then be drifted under a large field of 10$^4$ V/cm to create avalanche amplification. The sensitivity of this detector reaches $\sim$10$^{-43}$ cm$^2$ for DM-electron scattering and $\sim$10$^{-42}$ cm$^2$ for DM-nucleon scattering in a day. 

In comparison with the current SuperCDMS and EDELWEISS experiments, there are two main differences: (1) charge creation and (2) internal amplification. For the former, in our technology, the charge is created by ionization of impurities, which allows an experiment to access even lower energy deposition ($\sim$0.1 eV) comparing to SuperCDMS or EDELWEISS at which the charge is mostly created by ionization of Ge ($\sim$50 eV). In the case of the latter, we propose to internally amplify charge using avalanche while SuperCDMS and EDELWEISS internally amplify signal through emission of Luke phonons.  

\section*{Acknowledgments}
 We would like to thank Christina Keller for a careful reading of this manuscript.  This work was supported in part NSF OIA 1434142, DOE grant DE-FG02-10ER46709, the Office of Research at the University of South Dakota and a research center supported by the State of South Dakota.

\end{document}